# Black phosphorus Q-switched and mode-locked Er:ZBLAN fiber lasers at 3.5 μm

Zhipeng Qin,[1] Ting Hai,[1] Guoqiang Xie,[1,*] Jingui Ma,[1] Peng Yuan,[1] Liejia Qian,[1] Lei Li,[2] Luming Zhao,[2] and Deyuan Shen[2]

[1]Key Laboratory for Laser Plasmas (Ministry of Education), Collaborative Innovation Center of IFSA (CICIFSA), School of Physics and Astronomy, Shanghai Jiao Tong University, Shanghai 200240, China

[2]Jiangsu Key Laboratory of Advanced Laser Materials and Devices, Jiangsu Collaborative Innovation Center of Advanced Laser Technology and Emerging Industry, School of Physics and Electronic Engineering, Jiangsu Normal University, Xuzhou 221116, China

**Abstract**
With the proposal of dual-wavelength pumping (DWP) scheme, DWP Er:ZBLAN fiber lasers at 3.5 μm have become a fascinating area of research. However, limited by the absence of suitable saturable absorber, passively Q-switched and mode-locked fiber lasers have not been realized in this spectral region. Based on the layer-dependent bandgap and excellent photoelectric characteristics of black phosphorus (BP), BP is a promising candidate for saturable absorber near 3.5 μm. Here, we fabricated a 3.5-μm saturable absorber mirror (SAM) by transferring liquid-phase exfoliated BP flakes onto a gold-coated mirror. With the as-prepared BP SAM, we realized stable Q-switching and continuous-wave mode-locking operations in the DWP Er:ZBLAN fiber lasers at 3.5 μm. To the best of our knowledge, it is the first time to achieve passively Q-switched and mode-locked pulses in 3.5 μm spectral region. The research results will not only promote the development of 3.5-μm pulsed fiber lasers but also open the photonic application of two-dimensional materials in this spectral region.

## 1. Introduction

The past decade has seen the rapid development of mid-infrared fiber lasers which is driven by the great demand in a wide range of applications such as spectroscopy, defense, trace gas sensing, and greenhouse gases monitoring, etc [1-4]. Especially, in recent years, 3.5-μm Er:ZBLAN fiber laser has attracted wide attention with the advance of fluoride fiber fabrication and the proposal of dual-wavelength pumping (DWP) scheme [5]. Great efforts have been made to investigate the dynamics of Er:ZBLAN fiber laser by numerical simulation method, resulting in the discoveries of energy transfer up-conversion ($^4F_{9/2}+^4I_{11/2}\rightarrow{}^4S_{3/2}+^4I_{13/2}$) and excited state absorption ($^4F_{9/2}\rightarrow{}^4F_{7/2}$) [6, 7]. Experimentally, a wide tuning range of 450 nm has been demonstrated in DWP grating-tuned Er:ZBLAN fiber laser [8]. Besides, a record output power of 5.6 W has been demonstrated in a 3.55-μm Er:ZBLAN all-fiber laser [9]. Experimental results have shown power scaling and wavelength tunability potential of the DWP Er:ZBLAN fiber lasers operating in the continuous-wave (CW) regime. However, there has been no report on pulsed Er:ZBLAN fiber lasers at 3.5 μm so far.

Pulsed lasers with high energy and high peak power are very important in specific applications such as laser surgery, supercontinuum generation, Raman frequency shift, and optical parametric process [10-13]. Due to lack of a practical saturable absorber (SA) in 3.5-μm spectral region, the development of 3.5-μm pulsed lasers has been hindered. Semiconductor saturable absorber mirror (SESAM) is a commonly used SA, however, the commercial InGaAs SESAM only work below 3.0 μm [14]. It needs to develop a mid-infrared SA which can be used beyond 3.0 μm. Two-dimensional (2D) materials, such as graphene, topological insulators (TIs), and black phosphorus (BP), have been widely used as mid-infrared optoelectronic devices below 3 μm (optical modulators [15-18], field-effect transistor [19], photodetector [20], and photovoltaics [21]) due to their smaller bandgap, ultrafast carrier dynamics, and easy fabrication. For graphene SA, the challenge is the low modulation depth of graphene [22], restraining its application in fiber lasers where large modulation depth is usually required. Another 2D material candidate, TI such as $Bi_2Te_3$ has gapless surface states and bulk bandgap of 0.35 eV (correspond to photon energy of 3.55-μm laser) [24]. However, the indirect bulk bandgap and complicated preparation process reduce the potential as mid-infrared SA above 3.0 μm. BP, as a newly-emerged direct-bandgap 2D material, had been experimentally demonstrated to possess nonlinear optical absorption property based on the Z-scan or the balanced synchronous twin-detector measurement in the spectral range from 800 nm to 2779 nm [25-28]. Compared with graphene, BP possesses a faster recovery time and larger modulation depth [26, 28]. With liquid exfoliation method or mechanical exfoliation method [25, 27], the fabricated BP SAs have been widely used in Q-switched and mode-locked lasers below 3.0 μm [28-33], showing the excellent saturable absorption property. Different from gapless graphene and TIs, BP has a layer-dependent bandgap. Its bandgap decreases with the increasing thickness due to the interlayer interactions, and down to ~0.3 eV for the bulk BP. Therefore, it is a very promising candidate for mid-infrared 3.5-μm SA.

Here we fabricated a reflection-type BP saturable absorber mirror (SAM), which could work at the wavelength of 3.5 μm. This kind of BP SAM was fabricated by transferring the liquid-phase exfoliated (LPE) BP flakes onto a gold-coated mirror. We measured the nonlinear optical response of the as-prepared BP SAM at 3.5 μm, revealing a modulation depth of 7.7% and a saturation fluence of 0.36 μJ/cm$^2$. By using the as-prepared BP SAM as SA in a Er:ZBLAN fiber laser, we realized the stable Q-switching and CW mode-locking operations. To the best of our knowledge, it is the first time to achieve the passively Q-switched and CW mode-locked fiber lasers beyond 3 μm wavelength. Our research results show that BP is an excellent mid-infrared SA for 3.5-μm Q-switched and mode-locked fiber lasers.

## 2. Fabrication and characterization of BP SAM

The bulk BP was provided by a commercial supplier (XFNANO, Ltd.) and stored in an inert gas-filled glass bottle. In order to obtain multi-layer BP flakes, we firstly ground the mixture of bulk BP (~ 5 mg) and acetone (5 mL) for 10 minutes in a mortar. Then, the dispersions were ultra-sonicated at 40 kHz and 300 W power for one hour. After sonication, the upper suspension was dropped onto a gold-coated mirror. Finally, the as-prepared BP SAM was dried at 60 °C for hours to remove the residual acetone.

To measure the thickness of the as-prepared BP flakes, we transferred the BP flakes onto a silica plate and performed the thickness measurement with the atomic force microscopy (AFM), as shown in Fig 1(a). The thicknesses of the selected samples at three different areas were 52 nm, 88 nm, and 140 nm, respectively. In consideration of 0.5 nm thickness for monolayer BP [19], the 52 nm thickness corresponds to 104 layers. According to the BP bandgap equation

Eg≈(1.7/$n^{0.73}$+0.3) eV (n is the number of layers) [34], the as-prepared BP flakes had a bandgap of less than 0.36 eV, corresponding to a photon energy of 3.45 μm, which means that the as-prepared BP flakes could operate above 3.45 μm. Unlike graphene, which is chemically inert in atmosphere, BP is easy to be oxidized upon exposure to air, resulting in the formation of an oxide layer on the surface. Consequently, the oxidization process tends to saturate after two days in atmosphere under the protection of oxide [35]. Here, we measured the P 2p core level spectrum of the BP SAM, which was exposed to air for four days. The oxidization of BP was confirmed by the P 2p core level spectrum (Fig. 1(b)) with P2, P3, and P4 peaks at 131.8 eV, 132.8 eV, and 134.5 eV, respectively. Due to the formation of the oxidized BP on the surface, the inner BP could be protected by the surface oxidation layer. Therefore, the P1 peaks at 130.2 eV and 131.0 eV originate from the inner unoxidized BP. In order to demonstrate the nonlinear optical response of the surface-oxidized BP flakes at mid-infrared, we measured the saturable absorption at 3.5 μm wavelength. In the measurement, the light source is a tunable mid-infrared optical parametric amplifier (OPA) with an average power of 70 mW, a pulse duration of 213 fs, and a repetition rate of 1 MHz at 3.5 μm wavelength. Figure 2 clearly shows the saturable absorption of the surface-oxidized BP at 3.5 μm wavelength. From the fitting curve, we can obtain the modulation depth of 7.7% and the saturation fluence of 0.36 μJ/cm$^2$. To our knowledge, it is the first time to experimentally demonstrate the saturable absorption property of the BP beyond 3 μm.

## 3. Experimental setup

Figure 3 shows the schematic of the BP Q-switched and mode-locked Er:ZBLAN fiber lasers. The simplified energy diagram (Inset of Fig. 3) presents the energy transfer processes. Under the pumping of 970 nm, a mass of electrons can accumulate at the metastable level of $^4I_{11/2}$ and form the "virtual ground state". The 970 nm pump light was provided by the commercial laser diode (BWT Beijing Ltd.) with a pigtail fiber, having a numerical aperture (NA) of 0.2 and core diameter of 105 μm. Under further pumping of 1973 nm, the population inversion was realized between upper laser level of $^4F_{9/2}$ and lower laser level of $^4I_{9/2}$. Here, the 1973 nm pump light was generated from a homemade Tm-doped fiber laser. It had a maximum output power of 10 W, delivered from a double-clad silica fiber with $NA_{core}$ of 0.15 and core diameter of 10 μm. The two pump beams were combined together with a dichroic mirror $M_1$ and then focused into the Er:ZBLAN gain fiber (FiberLabs Inc.). The 2.8 m long double-clad Er:ZBLAN fiber had a core diameter of 16.9 μm with $NA_{core}$ of 0.15 which ensured not only the single-transverse-mode operation near 3.5 μm wavelength but also the high coupling efficiency of 1973 nm core pumping. For 970 nm pump, it was coupled into the circular inner clad with a clad diameter of 250 μm and $NA_{clad}$ of >0.52. For 1973 nm pump, it should be coupled into the core. In fact, we observed a small part of 1973 nm pump power coupled into inner clad. At the end of fiber, both the 970 nm and 1973 nm pump power, propagating in inner clad, were removed from the fiber using a clad-mode stripper (CMS). Only a little residual pump power, propagating in the fiber core, were output from the 8°-cleaved fiber facet. They were collimated together with the 3.5-μm laser by using a gold-coated off-axis parabolic mirror with an effective focal length of 12.7 mm. In order to avoid adverse effect on BP SAM, we employed a 45°-placed mirror $M_3$ to filter residual pump powers of 970 nm and 1973 nm. The mid-infrared laser was focused onto BP SAM with a highly antireflective-coated black diamond aspheric lens (f=4 mm).

## 4. Q-switching operation of Er:ZBLAN fiber laser with BP SAM

In the LPE BP dispersion, each flake differed from each other in thickness and flatness. Only the BP flake above 52 nm thickness could meet the requirement of saturable absorption at 3.5 µm. So, in the experiment, we need to transversally move the BP SAM to realize Q-switching. With the fixed 970 nm pump power of 1.76 W and 1973 nm pump power above 1.71 W, the Q-switched pulse train could be observed. The pulse train was captured by a mercury cadmium telluride (MCT) detector (PCI-9, VIGO System) and showed on a 1-GHz bandwidth digital oscilloscope (MDO 3102, Tektronix). Figure 4(a) presents three typical pulse trains at different average output powers of 71.4 mW, 88.5 mW, and 113.4 mW, respectively. The pulse intensity fluctuation ($\Delta_{RMS}$) was less than 4.2%, which means that the BP Q-switched pulses were stable. The pulse profiles are shown in Fig. 4(b). At an average output power of 113.4 mW, the Q-switched pulses had a pulse width of 2.09 µs. Compared with the pulse width of ~1.18 µs in BP Q-switched Er:ZBLAN fiber laser at 2.8 µm [28], the Q-switched pulses had a longer pulse width at 3.5 um. We attributed the longer pulse width to a smaller stimulated cross-section of Er:ZBLAN fiber at 3.5 µm [8]. Thus, more round trips are needed to exhaust the population in upper laser level.

Figure 5 shows the optical spectrum of BP Q-switched Er:ZBLAN fiber laser. The Q-switched pulse spectrum is around 3462 nm, measured with a mid-infrared spectral analyzer with a resolution of 0.22 nm (SIR5000, Ocean Optics). In this spectral region, there was no commercial SESAM for pulsed laser generation. The narrow-bandgap bulk BP fully displays its advantage as mid-infrared SA beyond 3 µm wavelength.

By increasing the launched pump power of 1973 nm, the evolution of average output power and pulse energy are presented in Fig. 6(a). As the launched 1973 nm pump power increased, the average output power increased linearly with a slope efficiency of 13.4% with respect to 1973 nm pump power. The maximum average power of 120 mW was obtained at the launched pump power of 2.1 W. The available maximum pulse energy was 1.8 µJ in the experiment. Figure 6(b) shows the repetition rate and pulse width versus the launched pump power of 1973 nm. The shortest pulse width of 2.05 µs was achieved at the repetition rate of 66.3 kHz.

## 5. CW mode-locking operation of Er:ZBLAN fiber laser with BP SAM

Nonlinear polarization rotation (NPR), one of the artificial SAs, has been usually employed in 3.0-µm mode-locked fluoride fiber lasers for femtosecond pulse generation [36, 37]. However, due to the small gain of fluoride fiber at 3.5 µm and large loss of optical elements in this spectral region, it is challenging to achieve mode-locking operation based on NPR at 3.5 µm. In order to realize mode-locking operation at 3.5 µm, we tried the as-prepared BP SAM to mode-lock 3.5 µm Er:ZBLAN fiber laser. Mode-locking operation was achieved under a lunched pump power of 2.4 W at 1973 nm. The mode-locked pulse train in Fig. 7(a) shows a pulse period of 36 ns, which is in consistent with the round-trip time of pulse in the cavity. The pulse train in millisecond time scale (Fig. 7(b)) indicates a CW mode-locking operation. We measured the radio frequency spectrum of the mode-locked pulses, showing a signal-to-noise ratio of 48 dB with a repetition rate of 22.77 MHz. The average output power of the mode-locked fiber laser was 60 mW. Limited by the low sensitivity of commercial autocorrelator at this wavelength and the low average power of the laser, regrettably, we failed to achieve the pulse duration by a commercial autocorrelator (APE pulse Check USB MIR). The mode-locked pulses had a spectral bandwidth of 1.4 nm at the central wavelength of 3462 nm (Fig. 7(d)), implying pulse duration with tens of picoseconds.

## 6. Conclusion

In conclusion, we demonstrated a mid-infrared 3.5-μm BP SAM and realized Q-switched and mode-locked operation at 3.5-μm wavelength for the first time. The experimental results show that BP SAM has an excellent saturable absorption performance at 3.5 μm, which is the first experimentally demonstrated SA beyond 3 μm wavelength. With the prepared BP SAM as SA, we successfully achieved Q-switched and mode-locked pulses from an Er:ZBLAN fiber laser at 3.5 μm. The Q-switched operation delivered an average power of 120 mW with a pulse energy of 1.83 μJ, a pulse width of 2.05 μs, and a repetition rate of 66.33 kHz at the wavelength of 3462 nm. In the CW mode-locking operation, we achieved picosecond pulses with an average power of 60 mW and a repetition rate of 27.8 MHz at 3462 nm. Our research results show that BP has a great potential as mid-infrared SA beyond 3μm wavelength, which is lacking currently in this wavelength region.

## Funding

The work is partially supported by the National Basic Research Program of China (Grant No. 2013CBA01505), the National Natural Science Foundation of China (Grant No. 61675130 and 11721091), the National Postdoctoral Program for Innovative Talents (No. 190375), and the Project funded by China Postdoctoral Science Foundation (No. 190375).

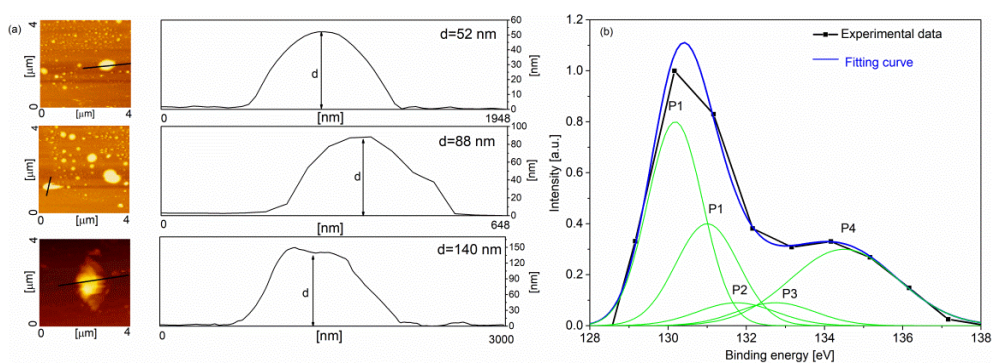

Fig. 1 (a) Two-dimensional morphologies of the BP flakes from three different areas on silica plate scanned by AFM and the corresponding height profiles of the section marked in AFM. (b) P 2p core level spectrum taken from BP flakes exposed in air after 4 days.

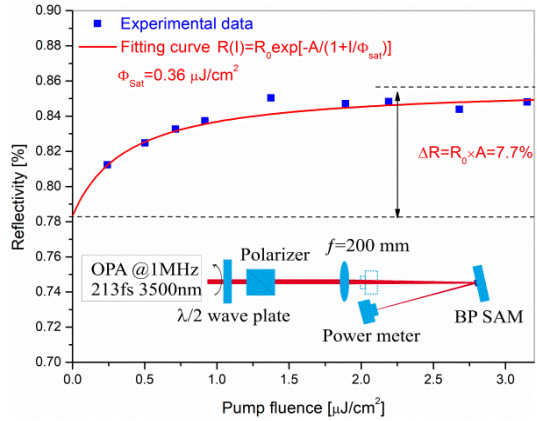

Fig. 2 Saturable absorption of BP SAM at 3.5 μm (upper) and the experimental measurement setup (lower).

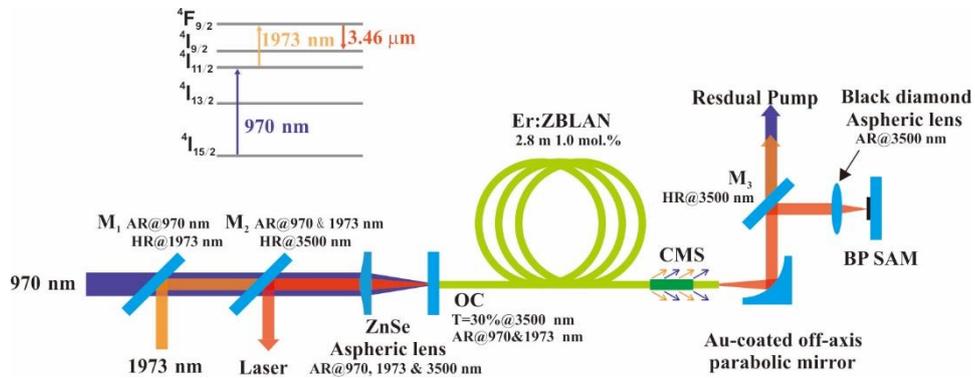

Fig. 3. Schematic of the BP Q-switched and mode-locked Er:ZBLAN fiber lasers. OC, output coupler; CMS, cladding mode stripper; BP SAM, black phosphorus saturable absorber mirror.

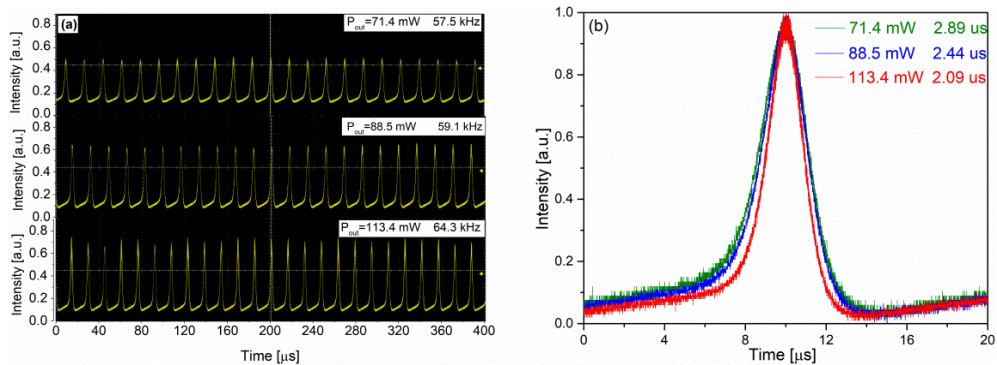

Fig. 4. (a) Q-switched pulse trains at the output powers of 71.4 mW, 88.5 mW, and 113.4 mW, respectively. (b) Corresponding pulse profiles.

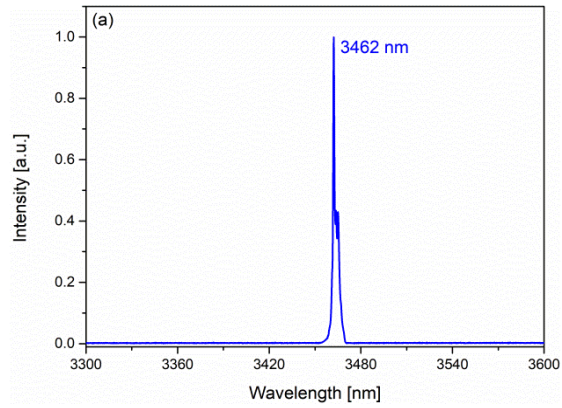

Fig. 5. Optical spectrum of the Q-switched pulses.

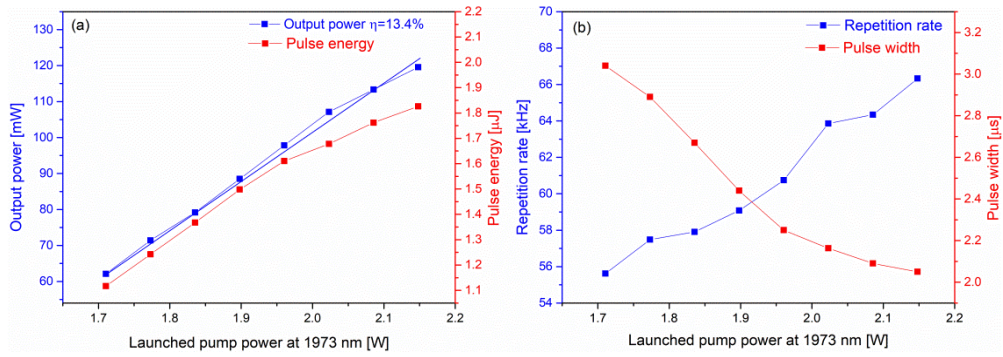

Fig. 6. (a) Average output power and pulse energy versus the launched pump power of 1973 nm. (b) Repetition rate and pulse width versus the launched pump power of 1973 nm.

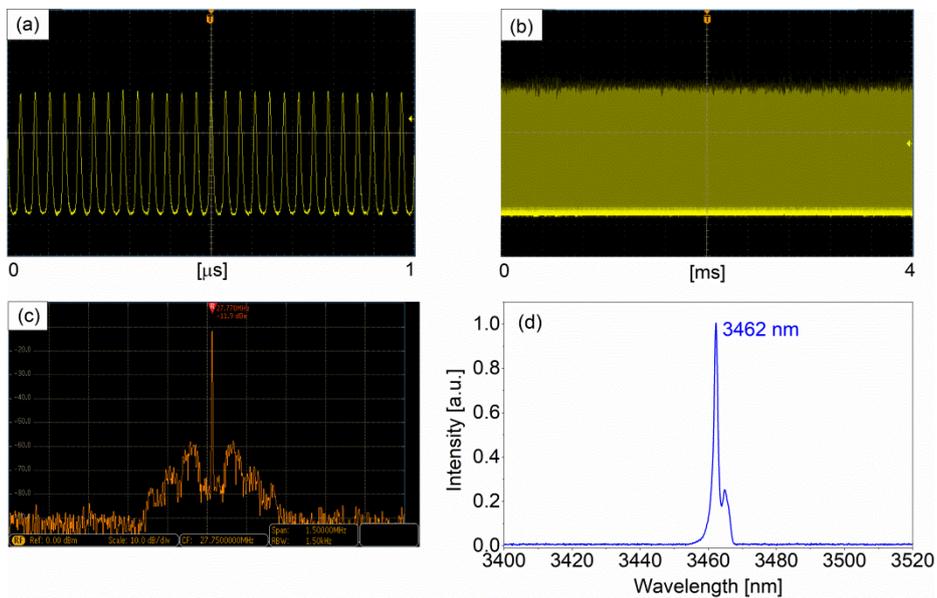

Fig. 7. (a) Mode-locked pulse train in microsecond time scale. (b) Mode-locked pulse train in millisecond time scale. (c) Radio frequency spectrum. (d) Mode-locked pulse spectrum.